\documentclass[10pt,twocolumn,english,aps,prl,nofootinbib,showkeys,preprintnumbers]{revtex4}
\usepackage[utf8]{inputenc}
\usepackage{graphicx}
\usepackage{amsmath}
\usepackage{amssymb}
\usepackage{amsfonts}
\usepackage{lscape}
\usepackage{hyperref}
\usepackage{url}
\usepackage{epsfig}
\usepackage{color}
\usepackage{bm}
\usepackage{tabularx}
\newcommand{\be}{\begin{equation}}
\newcommand{\ee}{\end{equation}}
\newcommand{\ba}{\begin{eqnarray}}
\newcommand{\ea}{\end{eqnarray}}

\def\ga{\mathrel{\raise.3ex\hbox{$>$\kern-.75em\lower1ex\hbox{$\sim$}}}}
\def\la{\mathrel{\raise.3ex\hbox{$<$\kern-.75em\lower1ex\hbox{$\sim$}}}}

\def\Msun{M_\odot}
\def\LQCD{\Lambda_{\rm QCD}}
\def\tQCD{t_{\rm QCD}}
\def\fPBH{f_{\rm PBH}}
\def\MPBH{M_{\rm PBH}}
\def\Gsph{\Gamma_{\rm sph}}
\def\Teff{T_{\rm eff}}
\def\Tth{T_{\rm th}}
\def\aW{\alpha_{\rm W}}
\def\dCP{\delta_{\rm CP}}
\def\nB{n_{\rm B}}
\def\OR{\Omega_{\rm R}}
\def\OM{\Omega_{\rm M}}
\def\OB{\Omega_{\rm B}}
\def\OPBH{\Omega_{\rm PBH}}
\def\ODM{\Omega_{\rm DM}}

\makeatother

\begin{document}

\title{A common origin for baryons and dark matter 
}

\author{Juan Garc\'ia-Bellido$^a$}
\email{juan.garciabellido@uam.es}

\author{Bernard Carr$^{b,c}$}
\email{B.J.Carr@qmul.ac.uk}

\author{S\'ebastien Clesse$^{d,e}$}
\email{sebastien.clesse@uclouvain.be \\ }

\affiliation{$^a$Instituto de F\'isica Te\'orica UAM-CSIC, Universidad Auton\'oma de Madrid,
Cantoblanco, 28049 Madrid, Spain\\
$^b$School of Physics and Astronomy, Queen Mary University of London, Mile End Road, London E1 4NS, UK\\
$^c$Research Center for the Early Universe, University of Tokyo, Tokyo 113-0033, Japan\\
$^d$Cosmology, Universe and Relativity at Louvain (CURL), Institut de Recherche en Mathematique et Physique (IRMP), Louvain University, 2 Chemin du Cyclotron, 1348 Louvain-la-Neuve, Belgium\\
$^e$Namur Institute of Complex Systems (naXys), University of Namur, Rempart de la Vierge 8, 5000 Namur, Belgium}

\date{\today}

\begin{abstract}
The origin of the baryon asymmetry of the Universe (BAU) and the nature of dark matter are two of the most challenging problems in cosmology. {We propose a scenario in which the gravitational collapse of large inhomogeneities at the quark-hadron epoch generates both the baryon asymmetry and dark matter in the form of primordial black holes (PBHs). This is due to the sudden drop in radiation pressure during the transition from a quark-gluon plasma to non-relativistic hadrons. The collapse to a PBH
%form black holes
 is induced by fluctuations of a light spectator scalar field in rare regions and is accompanied by  the violent expulsion of surrounding material, which might be regarded as a sort of ``primordial supernova" .}
%in only a few and far between domains.}
%We propose a scenario in which the gravitational collapse of pre-existing inhomogeneities at the quark-hadron epoch, enhanced by a transient sound-speed reduction then, generates both a baryon excess and dark matter in the form of primordial black holes (PBHs). The rare collapsing regions are produced by fluctuations of a light spectator scalar field, such as the axion, and generate a violent process that might be regarded as a ``primordial supernova". 
The acceleration of protons to relativistic speeds provides the ingredients for efficient baryogenesis around the collapsing regions and its subsequent propagation to the rest of the Universe. This scenario naturally explains why the observed BAU is of order the PBH collapse fraction and why the baryons and dark matter have comparable densities. The predicted PBH mass distribution ranges from sub-solar to several hundred solar masses. This is compatible with current observational constraints and could explain the rate, mass and low spin of the black hole mergers detected by LIGO-Virgo. Future observations will soon be able to test this scenario. 
\end{abstract}


\maketitle

{\em Introduction}. The first LIGO-Virgo detection~\cite{LIGO1st} of gravitational waves from the coalescence of two very massive black holes has triggered renewed interest in primordial black holes (PBHs) as dark matter (DM)~\cite{PBHDM}. Their abundance and mass distribution has intrigued both cosmologists and particle physicists~\cite{Rev}. If the PBHs were generated in the early radiation-dominated Universe from the gravitational collapse of large curvature fluctuations,  then they would have formed shortly after falling within the Hubble horizon with a mass
\be
\MPBH \simeq 0.5 \,  \gamma \,  g_*(T)^{-1/2} (T/{\rm GeV})^{-2}  M_\odot  \, . 
\ee
Here $\gamma$ is the fraction of the Hubble horizon mass ending up in the black hole
 %(around $0.2$ in a simple analysis~\cite{carr1975}
{(with $\gamma \la 1$  in general and $\gamma \approx 0.2$ in a simplified analysis~\cite{carr1975})},
%{ and $\gamma \lesssim 1$  in general}), 
$T$ is the temperature of the background Universe and $g_*(T)$ is the number of degrees of freedom then. $\MPBH$ is of order the Chandrasekhar mass, $M_{\rm Ch} \approx 1.4\,\Msun$, for PBHs forming at the Quantum Chromodynamics (QCD) scale, $\Lambda_{\rm QCD}\approx 200$ MeV. At this temperature, quarks and gluons form baryons (protons and neutrons) and mesons (pions) and the number of relativistic degrees of freedom drops abruptly. Also the sound speed dips, exponentially enhancing the collapse probability for  any large curvature fluctuation that enter the horizon then~\cite{PBHQCD}. The fraction of domains undergoing collapse is necessarily tiny, even if the PBHs provide all the DM. However, because they are non-relativistic, their density dilutes more slowly than the surrounding radiation until they dominate the expansion of the Universe at matter-radiation equality.

The sudden gravitational collapse of the mass within the Hubble horizon at the QCD epoch releases a large amount of entropy and generates a relativistically expanding shock-wave, with an effective temperature well above that of the surrounding plasma. Such high density {\em hot spots} might be {regarded as} 
%likened to 
\textit{primordial supernovae} and provide the out-of-equilibrium conditions required to generate a baryon asymmetry through the well-known electroweak sphaleron transitions responsible for Higgs windings around the electroweak (EW) vacuum~\cite{Asaka:2003}.  In this process, the charge-parity (CP) symmetry violation of the standard model of particle physics suffices to generate a local baryon-to-photon ratio of order one or larger. The hot spots are separated by many horizon scales {(thousands of kilometers) at the time of formation,}
 %{(by thousands of kilometers)} 
%and 
{while}
there is initially no matter-antimatter asymmetry in the rest of the Universe. However, since the baryons are relativistic {at formation}, they propagate away from the hot spots at the speed of light and become homogeneously distributed well before big bang nucleosynthesis.  The large initial local baryon asymmetry is thus diluted to the tiny observed global BAU.

The ratio of the energy densities of matter and radiation (relativistic species) at any time is 
\be
\frac{\OM}{\OR} = \frac{\OB+\ODM}{\OR} \simeq \frac{1700}{g_*(z)}\,\frac{1+\chi}{1+z}\,,
\ee
where $\chi \equiv \ODM/\OB \approx 5$ is the ratio of the DM and baryonic densities. At PBH formation, the fraction of domains that collapse is
\be
\beta = \frac{\OPBH}{\OR} = \fPBH\,\frac{\chi\,\OB}{\OR} \simeq
\fPBH\,\frac{\chi\,\eta}{\,g_*(T)} \frac{\rm 0.7\,GeV}{T}\,,
\ee
where $\fPBH \equiv \OPBH/\ODM$ is the fraction of the DM in PBHs and $\eta=n_{\rm B}/n_\gamma=6\times10^{-10}$ is the observed BAU. Therefore, for PBH formation at the QCD epoch, we have $\beta\sim\eta\sim10^{-9}$ if PBHs constitute {\em all} the DM. This relationship suggests that baryogenesis is somehow linked with PBH formation and that the smallness of the BAU reflects the rarity of the Hubble domains that collapse. Here we present a brief outline of a scenario with these features and derive the expected PBH mass distribution.  A more detailed description of our proposal - including the mechanism for generating curvature fluctuations via a spectator field and the various fine-tunings involved - can be found in a companion paper~\cite{Carr:2019hud}. 

{\em The quark-hadron transition}. In order for PBHs to form at the QCD epoch one needs large curvature fluctuations to  enter the horizon at the right time for relativistic particles to undergo gravitational collapse. One might fine-tune the inflationary dynamics (e.g. using the late plateau arising in critical Higgs inflation~\cite{CHI}) to produce a peak in the power spectrum of curvature fluctuations at the solar-mass scale with an amplitude several orders of magnitude larger than at the CMB scale~\cite{CGB:2015}. Large non-Gaussianity might further enhance the probability of gravitational collapse~\cite{Ezquiaga:2018}. However, such a fine-tuned peak is not required in our scenario because the sound speed drops abruptly by 30\% during the QCD transition due to the creation of non-relativistic protons and neutrons from quarks and gluons~\cite{cs2QCD}. This means that the radiation pressure, which usually prevents the collapse of mild inhomogeneities, suddenly drops, lowering the critical curvature $\zeta_c$ needed for PBH formation. Since the probability of collapse is exponentially sensitive to $\zeta_c$~\cite{carr1975}, they can form more easily. We need just a billionth of the domains to collapse {to PBHs} to explain all the DM.  {As explained later, this condition could be met  in our Universe without enhancing the power spectrum of curvature fluctuations or any other parameter fine-tuning. }

{\em Electroweak baryogenesis at the QCD epoch}.  The gravitational collapse at the QCD epoch of an horizon-sized  ball of radiation into a solar-mass black hole would be an extremely violent process, with  particles acquiring energies a thousand times their rest mass from the gravitational potential energy released by  the collapse.  As shown by simulations of PBH formation in spherical symmetry by Musco et al.~\cite{musco}, energy and momentum conservation imply that particles which  do not fall into the black hole are driven out as a shock-wave towards the surrounding plasma. This is similar to the shock-wave that ejects the outer layers of a star when it explodes as a supernova, except that  the surrounding plasma is much denser in the early Universe context, allowing higher energy interactions.  In particular, the effective temperature of the ``hot spot" is above that of EW sphaleron transitions, inducing local windings of the Higgs field around the EW vacuum.  Through the chiral anomaly, such topological configurations are equivalent to the creation of baryon number~\cite{MS:2000}. Since the surrounding plasma (initially beyond the Hubble domain that collapsed to form a PBH) is much cooler, and the far-from-equilibrium conditions ensure that further sphaleron transitions cannot wash out the local baryon asymmetry. 

This means that all of the Sakharov conditions~\cite{sak} for producing the matter-antimatter asymmetry are met. However, the asymmetry generated can be much larger than in the usual cosmological scenario. This is because the effective CP violation in the standard model (SM) is strongly temperature-dependent ($\delta_{\rm CP} \propto T^{-12}$), and the amount coming from the CKM matrix \cite{CKM} is enough for the local baryon-to-photon ratio to exceed one.  Subsequently the impulse of the shock-wave will drive baryons from the hot spot around each PBH to the rest of the Universe, thereby diluting the global baryon-to-photon ratio to the observed value, $\eta \sim 10^{-9}$.

Let us estimate the energy available for the process of hot spot electroweak baryogenesis (HSEWB). Energy conservation implies that the change in kinetic energy due to the collapse of matter within the Hubble radius, $d_H$, down to the Schwarzschild radius of the PBH, $R_S = 2G\MPBH/c^2 = \gamma\,d_H$, is
\be
\Delta K \simeq \left(\frac{1}{\gamma} - 1\right)M_{\rm H} = \left( \frac{1-\gamma}{\gamma^2} \right) \,\MPBH \, .
\ee
Note that the smaller the value of $\gamma$, the more compact the resulting PBH and the larger the kinetic energy of ejected particles. To estimate the energy acquired per proton $E_0$ in the expanding shell, we note that the number density of protons between the QCD transition and proton freeze-out ($20 \,  {\rm MeV} < T < 200 \, {\rm MeV}$) is that of a non-relativistic species, 
\be
n_{\rm p}(x) = 1.59\times10^{40}\,x^{-3/2}\,e^{-x}\,{\rm cm}^{-3}  \, ,
\ee 
with $x \equiv m_{\rm p}/T$. Therefore
\be
E_0 = \frac{\Delta K}{n_{\rm p}\,\Delta V} \simeq  
100\ g_*(x)\,x^{-5/2}\,e^{x}\,{\rm GeV}\, ,
\ee
where $\Delta V \equiv V_{\rm H} - V_{\rm PBH}$ is the difference between the Hubble and PBH volumes. We have used $\gamma=0.2$ as a conservative estimate but note that $E_0$ scales as 
%{$(1 - \gamma) / \gamma^2$}.
{$(\gamma + \gamma^2 +\gamma^3)^{-1}$. }
At the same time, the density of the relativistic plasma surrounding the collapse horizon is huge, $n_{\rm gas}(x) = 1.64\times10^{41}\, x^{-3}\,{\rm cm}^{-3}$, so it behaves like a wall for the escaping relativistic protons.

For a  PBH formed at $T\approx 140 \, {\rm MeV}$, the energy released and thus effective temperature is given by
\be
\Delta K = \frac{3}{2} N_{\rm p} k_{\rm B}\Teff  \ \ \Rightarrow\ \ k_{\rm B}\Teff = \frac{2}{3} E_0 \simeq {5 \, {\rm TeV}} \, ,
\ee
which is well above the sphaleron barrier and thus the sphaleron transition rate per unit volume at this temperature is $\Gsph\sim\aW^4\,\Teff^4$~\cite{MS:2000}. 
%The baryon asymmetry induced by the ultra-relativistic partons heating up the surrounding plasma is then~\cite{Asaka:2003}
{The ultra-relativistic partons (here mainly protons) produce jets that heat up the surrounding plasma and induce a baryon asymmetry~\cite{Asaka:2003}}
\be\label{nbs}
\eta \simeq \frac{7\nB}{s} \simeq \frac{7n_{\rm par}}{s}\times\Gsph(\Teff)\,V_{\rm H} \,\Delta t\times \dCP \,,
\ee
where $n_{\rm par}$ is the number density of the partons (here protons and antiprotons), $\Delta t\sim2\times10^{-5}\,{\rm s}\,(200\,{\rm MeV}/T)^2$ is the duration of the sphaleron process and the standard model CP violation parameter is~\cite{MS:2000} 
\be
\dCP(T) = 3\times10^{-5}\left(20.4\,{\rm GeV}/T\right)^{12} \, .
\ee
The entropy density in  the thermalized plasma surrounding each PBH is $s = (2\pi^2/45) \,g_{*S}\,\Tth^3$ at temperatures $\Tth\ll\Teff$; this quenches the sphaleron transitions and prevents baryon washout. The production of baryons is thus very efficient for $x\ga5$, giving  $\nB \ga n_\gamma$ or $\eta \ga 1$ locally. Note, however, that one cannot produce significantly more baryons than photons since they are soon brought into equilibrium with the rest of the plasma via standard model interactions. The dynamical process is actually rather complicated~\cite{Kurkela:2011} and will require further investigation.

This maximal BAU is then diluted as the protons propagate from the hot spots to the rest of the Universe. If the PBHs provide all the dark matter ($\fPBH=1$), one requires $\beta\sim10^{-9}$, and the distance between hot spots is then  $d\sim \beta^{-1/3}\,d_H(\tQCD)\sim 3000$~km, or 0.01 light-seconds. Moving at the speed of light, protons uniformly distribute the original baryon asymmetry to the rest of the Universe well before primordial nucleosynthesis ($t_{\rm BBN}\sim1-180$ s), thus diluting the initial baryon asymmetry and explaining the relation $\eta\sim\beta$. 

\begin{figure*}[ht]
\centering
%\vspace*{5mm}
\includegraphics[width = 0.9\textwidth]{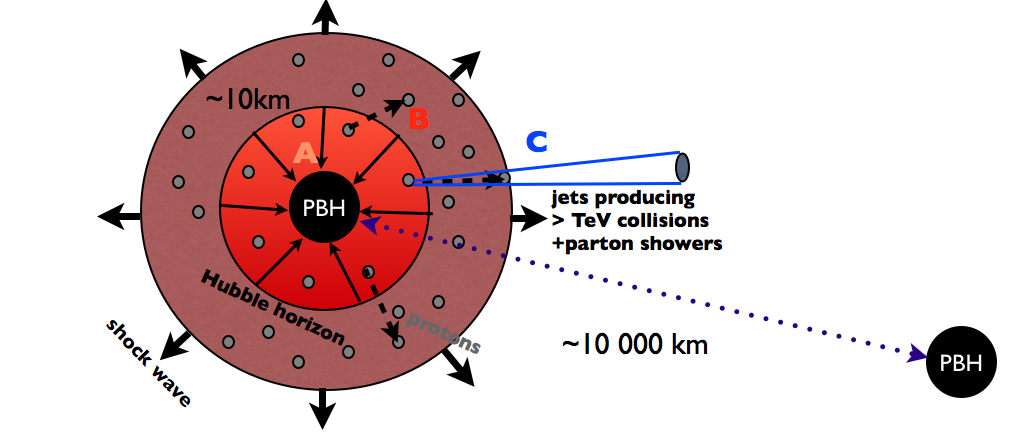}
%\vspace*{-3cm}
\caption{Qualitative representation of the three steps in our scenario. (A) Gravitational collapse to a PBH of the  curvature fluctuation at horizon re-entry. (B) Sphaleron transition in  hot spot around the PBH, producing $\eta \gtrsim \mathcal O(1)$ locally through EW baryogenesis. (C) Propagation of baryons to rest of Universe through jets, resulting in the observed BAU with $\eta \sim 10^{-9}$.  }
% {\bf The Hot Spot is *not* the shock wave. I would remove the words Hot Spot here.}}
\label{fig1}
\end{figure*}

{The DM-to-baryon ratio, $\chi\sim5$, {can also be} 
%is also 
explained in this scenario: most of PBHs are formed during or after the sudden drop of the sound-speed during the QCD transition, 
%when the temperature is low enough to produce a strong baryon asymmetry.  $\chi$ is thus given 
when the parton energies are high enough 
to produce a strong baryon asymmetry.  $\chi$ is thus given 
by the ratio of the black hole mass and the ejected mass, which is $\chi \approx \gamma / (1-\gamma) \approx 5$ if  $\gamma \approx 0.8$. 
 Lower values of $\gamma$ could nevertheless be accommodated if the temperature below which protons acquire enough energy to drive the baryon-producing sphaleron transitions is reduced, $T \la 100 \, {\rm MeV}$, so that only the massive PBHs formed at later time contribute to the BAU.  }
%only protons generated at later times 
%{($T \la 100 \, {\rm MeV}$)}
%($T \lesssim 200 \, {\rm MeV}$) 
%acquire enough energy to drive the baryon-producing sphaleron transitions. 
%{So only subdominant PBHs with a mass above approximately $10 \, \gamma  \, \Msun$ are able to drive the baryogenesis, and the ratio $\chi$ can estimated as the integral over the PBH mass distribution above this value, multiplied by $(1-\gamma)/\gamma$.  }
%The fraction {can be estimated from} 
%is just the 
%integral over the PBH mass distribution {above $ \MPBH \gtrsim 10 \, \gamma  \, \Msun$, divided by $\gamma$. }
%{\bf We need to explain this better.}
% from  $\MPBH\simeq1- 100\,\Msun$ divided by the integral over  the full range. 
The scenario is represented qualitatively in Fig.~\ref{fig1}.

{\em The origin of the large curvature fluctuations}. The softening of the equation of state during the quark-hadron transition boosts the formation of stellar-mass PBHs but does not alleviate the need for large curvature fluctuations. We propose that before or during the QCD epoch, a light stochastic spectator field~\cite{Hardwick:2017} induces {in rare} regions {an extra 
%peak in the 
curvature fluctuation},
% fluctuation above the  threshold} , 
{above the threshold required for PBH formation. }
% amplitude}, thus providing the critical amplitude 
%required for PBH formation.   
The spectator field is a curvaton; its quantum fluctuations during inflation permeate all space but its energy density is subdominant during both inflation and  the period after reheating. This field remains frozen during the radiation era ($m\ll H$) until its potential energy density (at the top of its potential) starts to dominate the total density of the Universe. At this point, the spectator field  in the still super-horizon regions triggers a second brief period of inflation, generating local non-linear curvature fluctuations which later reenter the horizon and collapse to form PBHs. In the rest of the Universe, {the field %directly fast 
rolls quickly towards the bottom of the potential and its fluctuations do not significantly impact the expansion.  This means that} the curvature fluctuations remain Gaussian, {at the same level as those observed in the CMB}, unaffected by the dynamics of the spectator field, and do not form PBHs.  {There are no isocurvature modes on cosmological scales, because the quantum fluctuations of both the inflaton and spectator fields scale with the Hubble rate during inflation, thereby correlating the large-scale curvature fluctuations with the PBH and baryon 
%density 
fluctuations. }

A natural candidate for the light spectator field is the QCD axion. Its existence is well-motivated, providing a robust solution to the strong CP problem. We assume that the associated Peccei-Quinn symmetry is spontaneously broken before inflation.  The axion potential at temperatures below a few GeV is  
\be
V(a) = m_a^{\rm eff}(T)^2\,f_a^2\,[1+\cos(a/f_a)] \, , 
\ee
where $m_a^{\rm eff}(T) = m_a\,(T/T_c)^{-7/2}$ for $T\ga T_c\sim 100$ MeV but  is constant and equal to the zero-temperature mass $m_a$ otherwise~\cite{axion}. For the QCD axion there is a relation between mass and decay constant, $m_a\,f_a \simeq  (75\,{\rm MeV})^2$.
Therefore, the axion will dominate the energy density of the Universe at temperatures below 
\be
T \approx  (60\,m_a^2\,f_a^2/\pi^2g_*)^{1/4} \approx 80 \, {\rm MeV} \, ,
\ee
but it already starts  rolling down the hill from the rms value generated during inflation, $a_{\rm ini} \ll f_a$, at $T\sim \rm{GeV}$.  

In most regions, this only marginally impacts the expansion rate, but in a few rare patches where the field lies exactly in the slow-roll region, it  produces a short period of  inflation until slow-roll ends at  $a_{\rm end} \simeq 8 \sqrt{\pi} f_a^2/{M}_{\rm P}$ where $M_{\rm P}$ is the Planck mass. The second inflationary period can last slightly more than one e-fold, which produces ${\cal O}(1)$ curvature fluctuations, according to the stochastic $\delta N$ formalism~\cite{Hardwick:2017fjo}.  The probability of collapse depends on the mean value of the axion (curvaton) field in our Hubble patch but it can be around $10^{-9}$, as required, if $f_a \gtrsim 10^{17} {\rm GeV}$.

\begin{figure*}[ht]
\centering
%\vspace*{5mm}
\includegraphics[width = 0.7\textwidth]{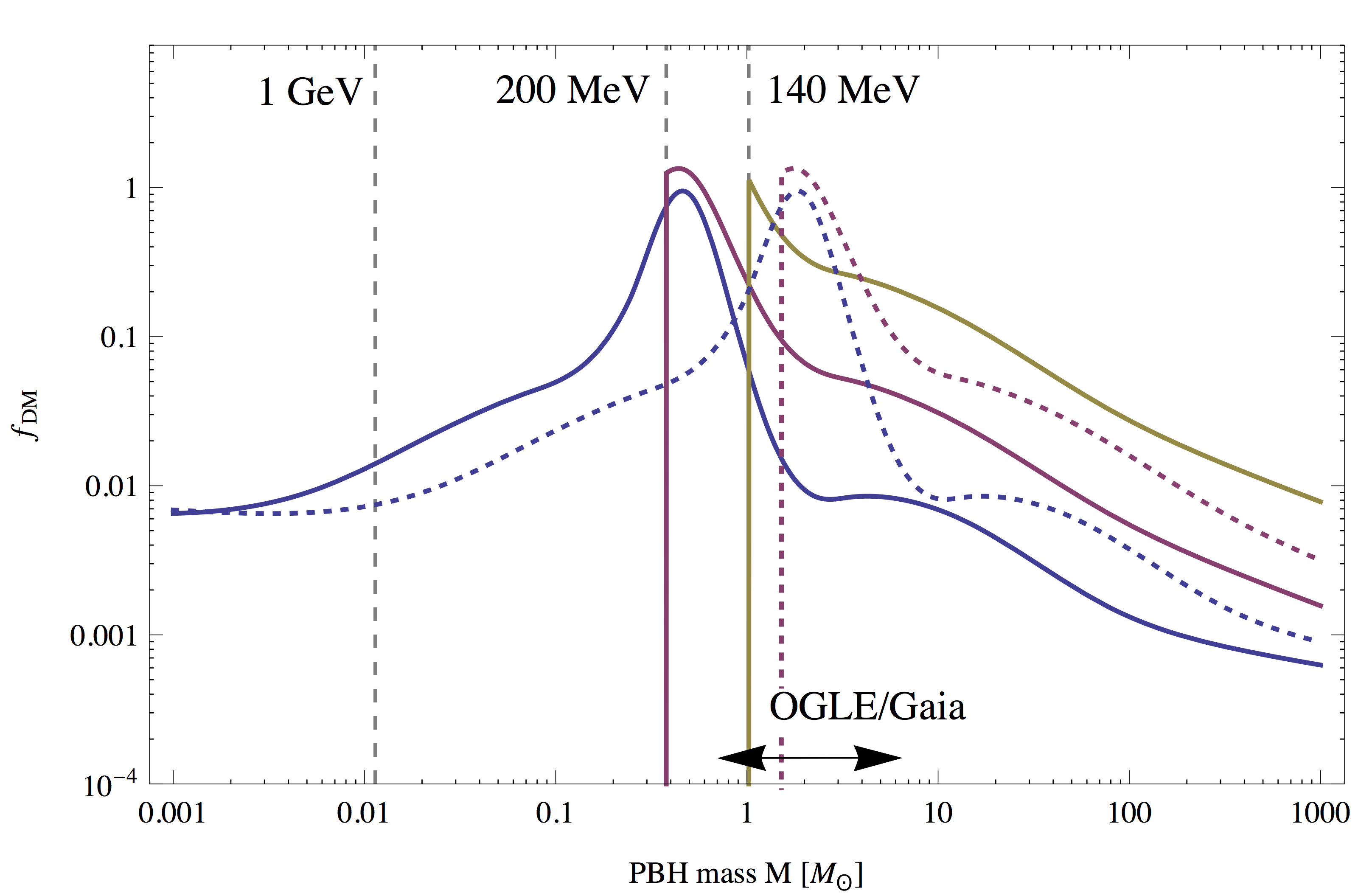}
%\vspace*{-3cm}
\caption{{PBH mass functions,  $f_{\rm PBH} \equiv ({\rm d} \rho_{\rm PBH} / {\rm d} \, \ln M )/ \rho_{\rm DM}$, for different curvaton models {and for a collapse efficiency $\gamma = 0.2$ (solid) and $\gamma = 0.8$ (dotted)}, the integrated PBH abundance correponding to the dark matter density in all cases. The vertical (grey) lines correspond to the temperature at PBH formation {(assuming $\gamma = 0.2$)}.  The top curves apply if the curvaton is identified with the QCD axion, with  $T_{\rm c} \simeq 200 \, {\rm MeV} $ {(red) and $T_{\rm c} \simeq 140 \, {\rm MeV} $ (yellow), assuming $f_{\rm a} = 0.15 \bar M_{\rm p}$, but the value of $f_{\rm a}$ only marginally impacts the shape of the distribution}.  
%For the axion model, 
In these cases PBH formation at $T \gtrsim T_{\rm c}$ is suppressed, which induces a sharp lower cut-off in the distribution.
 %(a sharp cut-off is assumed here).
 The blue curve applies for a more general curvaton field.   {The double arrow indicates 
%the position of 
the peak in the number of dark lenses from combined OGLE and Gaia microlensing observations~\cite{Wyrzykowski:2019}.  }}   }
\label{fig:fDM}
\end{figure*}

{\em The PBH mass distribution.} This is shown in Fig.~\ref{fig:fDM} and is a concrete prediction of our scenario. In the general curvaton case, shown by the lower curves {for $\gamma =0.2$ (solid lines) and $\gamma = 0.8$ (dotted lines)}, the largest density is associated with the horizon mass when protons become non-relativistic at $T\sim\LQCD$ and we have seen that this is of order the Chandrasekhar mass ($1.4 \, \Msun$). Then there is a small plateau associated with the temperature $T\sim m_\pi$ at which pions become non-relativistic. This also slightly changes the sound-speed and corresponds to $M \sim 30\,\Msun$, which may explain why LIGO-Virgo find so many black holes with that mass. At later times, the relativistic degrees of freedom again dominate the expansion of the Universe, so the PBH mass distribution declines quickly at larger masses, evading all the present constraints.
 
If the spectator field is the QCD axion, since its mass turns on abruptly, {the PBH production is strongly suppressed}
%there can be no light PBH production 
at temperatures above $T_{\rm c}$. Therefore the PBH mass distribution is naturally cut-off 
{on sub-solar masses},
%below around $1 \, \Msun$,
 as shown by the top curves in Fig.~\ref{fig:fDM}. This could explain the observed lack of subsolar microlenses{, whose significance is still debated~\cite{microlensing}}. 
In both cases, the majority of PBHs are in the {$0.1-10 \, M_\odot$} range {and no more than a few percent of the DM density is made of heavier PBHs.  Such a distribution 
%thus 
passes the various constraints on the abundance of massive PBHs~\cite{Carr:2016drx}}. 
%Moreover, 
The second peak in the distribution might explain the mass, rates and low spins of the black hole mergers detected so far by LIGO-Virgo. This is very different than the distribution expected for stellar black holes, which should exhibit a gap in the range $2 - 5 \, \Msun$ and be suppressed above $80 \, M_\odot$~\cite{Belczynski:2016}.  It is intriguing that an {excess} of {dark} microlensing events in this mass range has recently been reported from OGLE and Gaia observations of the Galactic bulge~\cite{Wyrzykowski:2019}.
 
In the near future, further LIGO-Virgo observations, upcoming microlensing and supernova lensing surveys, and a series of other electromagnetic probes~\cite{Ali-Haimoud:2019} should determine the mass spectrum of coalescing black holes~\cite{GWTC1} sufficiently well to test our scenario. In particular, LIGO-Virgo might confirm both the ``proton" peak and the ``pion" plateau at tens of solar masses.

{\em Addressing the fine-tunings.}   Our scenario naturally links the PBH abundance  to the baryon abundance and the BAU to the PBH collapse fraction ($\eta \sim \beta$). The spectator field mechanism for producing the required curvature fluctuations also avoids the need for a fine-tuned peak in the power spectrum, which has long been considered a major drawback of PBH scenarios.  One still needs fine-tuning of the mean field value to produce the observed values of $\eta$ and $ \beta$ (i.e. {$\sim 10^{-9}$}). However,  the stochasticity of the field during inflation (if it lasted for more than 60 e-folds) ensures that Hubble volumes exist with all  possible field values and this means that one can  explain the fine-tuning by invoking a single anthropic selection argument. 

{The argument is discussed in Ref.~\cite{Carr:2019hud} and depends on the fact that} only a small fraction of patches will have the PBH and baryon abundance required for galaxies to form.  
%In most of them, 
{In most others}, the field is too far from the slow-roll region to produce either PBHs or baryons.  Such patches lead  to radiation universes without any DM or matter-antimatter asymmetry.  In other (much rarer) patches, PBHs are produced too copiously, leading to rapid accretion of most of the baryons, as might have happened in ultra-faint-dwarf galaxies. This anthropic selection effect may therefore explain the observed value of $\eta$ and $\beta$. The connection between the rareness of the PBHs, responsible for later matter-domination, and subsequent structure formation is an important feature of our scenario.

{\em Conclusions}. It is well known that the early Universe can be used as a probe of fundamental physics at very high energies. The production of the BAU through CP-violating processes is one example of this, the usual assumption being that new high-energy physics generates the baryon asymmetry everywhere simultaneously via out-of-equilibrium particle decays or first-order phase transitions. However, in our scenario, the BAU is generated in local hot spots  through  the violent process of PBH formation at  the QCD transition, this being triggered by the sudden drop in the radiation pressure and  the presence of large amplitude curvature fluctuations. The only CP violation needed is that of the Standard Model and the same regions which generate the baryon asymmetry also produce PBHs with a density comparable to that of the baryons. 

  {A full analysis of the non-linear dynamics of gravitational collapse and out-of-equilbrium baryogenesis will require detailed numerical simulations. Future particle physics experiments with ultra-high-density heavy ion collisions in the 100 TeV range~\cite{Ellis:2016}  may be able to explore the high-energy sphaleron transitions  invoked by our proposal. Note that our model does not preclude some baryogenesis occuring at an  earlier epoch, providing the associated value of $\eta$ is much less than $10^{-9}$. }  

 {All the dark matter in our propoisal  is made of PBHs with a mass distribution which peaks around a solar mass. 
 %is such that it 
This passes the current observational 
 %astrophysical and cosmological 
 constraints on the PBH abundance, once the large uncertainties on lensing constraints are taken into account.   
 % it could have been recently detected through microlensing by the OGLE survey.  
However, accurate predictions of the PBH mass function will require numerical investigations of the stochastic dynamics of the curvaton, both during and after inflation, for different spectator fields. If LIGO-Virgo interferometers over the next few years can determine the mass distribution of the coalescing black holes, this will allow a comparison with the predictions of  Fig.~\ref{fig:fDM}.}

{\em Acknowledgements}. The authors thank Misha Shaposhnikov, Chris Byrnes, {Karsten Jedamzik, Jane MacGibbon}, Ilia Musco and Ester Ruiz Morales for useful comments and suggestions. JGB acknowledges support from the Research Project FPA2015-68048-03-3P [MINECO-FEDER] and the Centro de Excelencia Severo Ochoa Program SEV-2016-0597.  BC thanks the Research Center for the Early Universe (RESCEU) at the University of Tokyo for hospitality received during this work. The work of SC is supported by the Belgian Fund for Research F.R.S.-FNRS.

\end{document}